\documentclass[aps,floats]{revtex4}
\usepackage{amsmath,amssymb}
\usepackage{graphicx,epsfig}
\usepackage[greek,english]{babel}
\begin{document}
\bibliographystyle {plain}

\def\oppropto{\mathop{\propto}} 
\def\opsimeq{\mathop{\simeq}}
\def\opoverderline{\mathop{\overline}}
\def\operarrow{\mathop{\longrightarrow}}
\def\opsim{\mathop{\sim}} 
\def\opmin{\mathop{\min}} 
\def\opmax{\mathop{\max}} 

\def\fig#1#2{\includegraphics[height=#1]{#2}}
\def\figx#1#2{\includegraphics[width=#1]{#2}}

%\newcommand{\fig}[2]{\epsfxsize=#1\epsfbox{#2}} \reversemarginpar 

%%%%%%%%%%%%%%%%%%%%%%%%%%%%%%%%%%%%%%%%%%%%%%%%%%%%%%%%%%%%%%%%%%%%%%%%%%%%
\title{ Multifractality of eigenstates in the delocalized non-ergodic phase \\
 of some random matrix models : Wigner-Weisskopf approach } 

%%%%%%%%%%%%%%%%%%%%%%%%%%%%%%%%%%%%%%%%%%%%%%%%%%%%%%%%%%%%%%%%%%%%%%%%%%%%

\author{ C\'ecile Monthus }
 \affiliation{Institut de Physique Th\'{e}orique, 
Universit\'e Paris Saclay, CNRS, CEA,
91191 Gif-sur-Yvette, France}

\begin{abstract}

The delocalized non-ergodic phase existing in some random $N \times N$ matrix models is analyzed via the Wigner-Weisskopf approximation for the dynamics from an initial site $j_0$. The main output of this approach is the inverse $\Gamma_{j_0}(N)$ of the characteristic time to leave the state $j_0$ that provides some broadening $\Gamma_{j_0}(N) $ for the weights of the eigenvectors. In this framework, the localized phase corresponds to the region where the broadening $\Gamma_{j_0}(N) $ is smaller in scaling than the level spacing $\Delta_{j_0}(N) \propto \frac{1}{N}$, while the delocalized non-ergodic phase corresponds to the region where the broadening $\Gamma_{j_0}(N) $ decays with $N$ but is bigger in scaling than the level spacing $\Delta_{j_0}(N) $. Then the number $\frac{\Gamma_{j_0}(N)}{\Delta_{j_0}(N)} $ of resonances grows only sub-extensively in $N$. This approach allows to recover the multifractal spectrum of the Generalized-Rosenzweig-Potter (GRP) Matrix model [V.E. Kravtsov, I.M. Khaymovich, E. Cuevas and M. Amini, New. J. Phys. 17, 122002 (2015)]. We then consider the L\'evy generalization of the GRP Matrix model, where the off-diagonal matrix elements are drawn with an heavy-tailed distribution of L\'evy index $1<\mu<2$ : the dynamics is then governed by a stretched exponential of exponent $\beta=\frac{2 (\mu-1)}{\mu}$ and the multifractal properties of eigenstates are explicitly computed.

\end{abstract}

\maketitle

\section{ Introduction }

The recent huge activity in the field of Many-Body-Localization (M.B.L.)  (see the recent reviews \cite{revue_huse,revue_altman,revue_vasseur,revue_imbrie} and references therein)
has renewed the interest into some subtle properties of various Anderson Localization models. In particular, the MBL-delocalized phase, which is usually
expected to be ergodic and to follow the Eigenstate Thermalization Hypothesis (E.T.H.) \cite{deutsch,srednicki,nature,mite,rigol},
has been found to display anomalously slow dynamical properties \cite{alet_dyn,luitz_anomalous,znidaric_j}
and to be nontrivial (see the recent review \cite{review_mblergo,review_rare} and references therein). Another possibility that has been raised is 
the existence of a delocalized non-ergodic phase, as discussed in \cite{grover,harrisMBL,c_entropy,c_mblrgeigen,santos}. 
This delocalized non-ergodic scenario is usually explained within the point of view that the MBL transition is somewhat similar
to an Anderson Localization transition in the Hilbert space of 'infinite dimensionality'
 where the size of the Hilbert space grows exponentially
with the volume \cite{levitov,gornyi_fock,vadim,us_mblaoki,gornyi}. 
As a consequence, this issue has motivated many recent works
 to confirm or to rule out the existence of a delocalized non-ergodic phase
in the short-ranged Anderson model
 on the Bethe lattice, either with boundaries \cite{us_cayley,mirlin_tik}
or without boundaries, where this issue remains extremely controversial,
since many recent papers contain completely opposite conclusions
 \cite{biroli_nonergo,luca,mirlin_ergo,altshuler_nonergo,lemarie,ioffe}.

Within Random Matrix Models, 
the question of the existence of a non-ergodic delocalized phase
has been actually raised more than twenty years ago 
by Cizeau and Bouchaud \cite{cizeau}
in their pioneering work on Random L\'evy Matrices,
that has attracted a lot of interest among physicists
 \cite{medina,birolitopeigen,burda,metz,majumdar,biroli_levy,c_levy} 
and mathematicians \cite{soshnikov,benarous,
auffinger_peche,auffinger_gui,bordenave2013,bordenave,benaych,peche}.
More recently, the Generalized-Rosenzweig-Porter model has been proposed
 as the simplest matrix model exhibiting a delocalized non-ergodic phase
with an explicit multifractal spectrum for eigenvectors in \cite{kravtsov_rosen}.
It has been then revisited from various points of view,
namely via the statistics of the local resolvent \cite{biroli_rosen},
via the super-symmetry approach \cite{ossipov_rosen}
 and via the self-consistent cavity equations \cite{ioffe}.
In this paper, our goal is to propose still another point of view
based on the Wigner-Weisskopf approximation for the dynamics :
this approach is applied to the usual Generalized-Rosenzweig-Porter (GRP) model,
as well as to some L\'evy generalization of the GRP model that we introduce
(it should be stressed that it is different from the usual L\'evy Matrix Model of Cizeau and Bouchaud \cite{cizeau} mentioned above).

The paper is organized as follows.
Section \ref{sec_model} contains the definition of the models.
In section \ref{sec_loc}, we describe how
the multifractal properties of the localized phase and of the critical point
can be obtained by the strong disorder perturbative expansion.
In section \ref{sec_dyn}, the dynamics from an initial site $j_0$ is analyzed via the Wigner-Weisskopf approximation
in order to obtain the weights of the eigenvectors in the delocalized non-ergodic phase.
This general framework is then applied to
 the Generalized-Rosenzweig-Porter model in section \ref{sec_rosen}
and to its L\'evy generalization in section \ref{sec_levy}.
Our conclusions are summarized in section \ref{sec_conclusion}.

\section{ Models and notations }

\label{sec_model}

In this paper, we focus on $N \times N$ symmetric matrix models,
where the diagonal matrix elements $H_{ii}$ are $O(1)$ random variables drawn
with some distribution $P_{diag}(H_{ii})$, while the off-diagonal 
matrix elements $H_{i<j}$ are rescaled with respect to the system size $N$ 
with some exponent $a$
\begin{eqnarray}
H_{i<j}  = \frac{v_{ij}}{N^a}
\label{offa}
\end{eqnarray}
where $v_{ij}$ are $O(1)$ random variables drawn with some symmetric probability distribution $p_{off}(v)=p_{off}(-v)$.

\subsection{Generalized-Rosenzweig-Porter (GRP) model }

The Generalized-Rosenzweig-Porter model introduced in \cite{kravtsov_rosen}
and revisited from various points of view \cite{biroli_rosen,ossipov_rosen,ioffe}
corresponds to the case where the variance is finite and can be chosen to be unity
\begin{eqnarray}
\overline{v_{ij}^2} = \int_{-\infty}^{+\infty} dv v^2 p_{off}(v) = 1
\label{rosenvar}
\end{eqnarray}
Then the eigenvalues of the matrix remain finite $O(1)$ in the region 
\begin{eqnarray}
a \geq \frac{1}{2}
\label{arosenvar}
\end{eqnarray}

\subsection{L\'evy version of the Generalized-Rosenzweig-Porter (L\'evy-GRP) model }

We will also consider the case where $v_{ij}$ is drawn with some heavy-tailed
distribution with $0<\mu<2$
\begin{eqnarray}
p_{off}(v_{ij}) = \frac{ \mu }{ 2  \vert v_{ij} \vert^{1+\mu} } 
\theta \left( \vert v_{ij} \vert \geq 1 \right)
\label{pofflevy}
\end{eqnarray}
so that the variance does not exist in contrast to the case of Eq. \ref{rosenvar}.

The probability distribution of off-diagonal elements reads (Eqs \ref{offa}
and \ref{pofflevy})
\begin{eqnarray}
P_{off}(H_{ij}) = \frac{ \mu }{ 2 N^{a\mu} \vert H_{ij} \vert^{1+\mu} } 
\theta \left( \vert H_{ij} \vert \geq N^{-a} \right)
\label{tailoff}
\end{eqnarray}
The typical value scales as expected as
\begin{eqnarray}
 H_{ij}^{typ} \propto N^{-a}
\label{offtyp}
\end{eqnarray}
but the maximum value seen by some given site $j_0$ is much bigger and scales as
\begin{eqnarray}
\opmax_{j \ne j_0} ( H_{j_0j} ) \propto N^{- \left(a- \frac{1}{\mu} \right)}
\label{offmax}
\end{eqnarray}
As a consequence, the eigenvalues of the matrix remain finite $O(1)$ in the region 
\begin{eqnarray}
a \geq \frac{1}{\mu}
\label{alevy}
\end{eqnarray}
that replaces Eq. \ref{arosenvar}.

\section{ Multifractal properties in the localized phase and at criticality }

\label{sec_loc}

\subsection{Strong Disorder perturbative expansion }

In the Strong Disorder perturbative expansion,
one considers the perturbation theory in the off-diagonal terms
 \cite{kravtsov_rosen,c_levy,us_prbm}.
At order $0$, the eigenvectors are completely localized on a single site
\begin{eqnarray}
  \vert \phi_j^{(0)} > = \vert j >
\label{phi0}
\end{eqnarray}
and the eigenvalues are given by the $O(1)$ diagonal matrix elements
\begin{eqnarray}
  E_j^{(0)}  = H_{jj}
\label{eigen0}
\end{eqnarray}

At first order in the off-diagonal elements that decay with the size $N$,
the eigenvalues remain unchanged
\begin{eqnarray}
  E_j^{(0+1)}  = H_{jj}
\label{eigen01}
\end{eqnarray}
 while the eigenstates become
\begin{eqnarray}
\vert \phi^{(0+1)}_j > = \vert j > +\sum_{k \ne j}  \frac{H_{kj} }{H_{jj}-H_{kk} }  \vert k >
\label{eigenper}
\end{eqnarray}
The idea is that this expression makes sense
 as long as the number of resonances defined by 
$\vert  H_{kj} \vert > \vert H_{jj}-H_{kk}  \vert$ does not grow with the system size $N$,
and this corresponds to the Localized phase.
The multifractal properties of the eigenstates 
 can be then derived from the weights of Eq. \ref{eigenper}
\begin{eqnarray}
w_j^{loc}(j_0)  && \equiv   \vert <j_0 \vert \phi^{(0+1)}_j > \vert^2 \simeq  \frac{  H_{jj_0}^2} {( H_{jj} - H_{j_0j_0} )^2  } 
\label{wjloc}
\end{eqnarray}

\subsection{ Multifractality in the Localized phase of the Generalized-Rosenzweig-Porter (GRP) model}

The typical value of the weights of Eq. \ref{wjloc}
corresponds to finite energy differences $H_{jj} - H_{j_0j_0} =O(1) $ 
\begin{eqnarray}
[w_j^{loc} ]_{typ} &&  \propto  N^{-2a} 
\label{wjloctyp}
\end{eqnarray}
while the maximal weight occurs for nearby states
separated by level spacing 
\begin{eqnarray}
\Delta_{j_0} (N) \equiv \vert H_{j_0j_0}-H_{next} \vert = \frac{1}{N \rho(H_{j_0j_0})}
\label{levelspacing}
\end{eqnarray}
and scales as
\begin{eqnarray}
[w_j^{loc} ]_{max} &&  \propto \frac{N^{-2a} }{\Delta^2_{j_0} (N)} \propto  N^{-2(a-1) } 
\label{wjlocmax}
\end{eqnarray}
This shows that the localized phase corresponds to the region \cite{kravtsov_rosen}
\begin{eqnarray}
a^{loc} > a_c=1
\label{alocrosen}
\end{eqnarray}

The probability distribution of the weight of Eq. \ref{wjloc} 
\begin{eqnarray}
{\cal P}^{loc}_N (w)  &&  = \int dH_{jj} P_{diag}(H_{jj}) \int dv p_{off}(v)
 \delta \left( w-  \frac{  N^{-2a} v^2 } {( H_{jj} - H_{j_0j_0} )^2  } \right)
\nonumber \\
&& =  \frac{ \int dv p_{off}(v) \vert v \vert \left[  P_{diag}\left(H_{j_0j_0} + v \frac{N^{-a}}{\sqrt{w} }  \right) + P_{diag}\left(H_{j_0j_0} - v \frac{N^{-a}}{\sqrt{w} }  \right) \right]}{ 2 N^a w^{\frac{3}{2}}}
\label{pwjloc}
\end{eqnarray}
displays the power-law tail
\begin{eqnarray}
{\cal P}^{loc}_N (w)  &&  \opsimeq_{w \to +\infty}  \frac{ P_{diag}\left(H_{j_0j_0}  \right) \int dv p_{off}(v) \vert v \vert }{  N^a w^{\frac{3}{2}}}
\label{pwjloctail}
\end{eqnarray}
For the exponent
\begin{eqnarray}
\alpha \equiv - \frac{ \ln w}{\ln N}
\label{alphaweight}
\end{eqnarray}
Eq. \ref{pwjloctail} translates into the multifractal spectrum for the probability
$\Pi^{loc} (\alpha) $ of $\alpha$
\begin{eqnarray}
\Pi^{loc} (\alpha)  &&  \simeq N^{ \frac{\alpha}{2} -a } 
\label{pchiloc}
\end{eqnarray}
The typical exponent corresponding to a finite probability $\Pi^{loc} (\alpha_{typ})  =O(1)$ is $\alpha_{typ}=2a$ in agreement with Eq. \ref{wjloctyp},
while the exponent associated to the maximal weight of Eq. \ref{wjlocmax}
 is $\alpha_{min}=2 (a-1)$ and corresponds to a probability of order
 $\Pi^{loc} (\alpha_{min}) \sim \frac{1}{N}$.
So the number ${\cal N}^{loc}(\alpha)$ of weights scaling as $w \propto N^{-\alpha}$ involves the linear multifractal spectrum \cite{kravtsov_rosen}
\begin{eqnarray}
{\cal N}^{loc}_{a>1} (\alpha)  &&  \simeq N^{ \frac{\alpha}{2} - (a-1)  }   \theta \left( 2 (a-1) \leq  \alpha \leq 2a \right)
\label{dqloc}
\end{eqnarray}

\subsection{ Multifractality in the Localized phase of the L\'evy-GRP model}

\label{sec_rosenloc}

The above calculation, in particular Eq. \ref{pwjloctail},
shows that the multifractal properties remain the same as long as the 
average of the absolute value of $v$ converges
\begin{eqnarray}
 \int dv p_{off}(v) \vert v \vert <+\infty
\label{cvabsv}
\end{eqnarray}
i.e. in the region $1<\mu<2$ of the L\'evy case of Eq. \ref{pofflevy},
leading to
\begin{eqnarray}
{\cal N}^{loc}_{1<\mu<2;a>1} (\alpha)  &&  \simeq N^{ \frac{\alpha}{2} - (a-1)  }   \theta \left( 2 (a-1) \leq  \alpha \leq 2a \right)
\label{dqloclevymubig}
\end{eqnarray}

For $0<\mu<1$ where Eq. \ref{cvabsv} diverges,
the probability distribution of the weight of Eq. \ref{wjloc} reads
using Eq. \ref{pofflevy}
\begin{eqnarray}
{\cal P}^{loc}_{0<\mu<1} (w)  &&  = \int dH_{jj} P_{diag}(H_{jj}) \int dv 
 \frac{ \mu }{ 2  \vert v \vert^{1+\mu} } 
\theta \left( \vert v \vert \geq 1 \right)
 \delta \left( w-  \frac{  N^{-2a} v^2 } {( H_{jj} - H_{j_0j_0} )^2  } \right)
\nonumber \\
&& = \frac{\mu}{2  N^{a \mu} w^{1+\frac{\mu}{2}}} \int dH_{jj} P_{diag}(H_{jj}) \vert H_{jj} - H_{j_0j_0}  \vert^{-\mu} \theta \left(  \vert H_{jj} - H_{j_0j_0} \vert \geq  \frac{  N^{-a}  }
 { \sqrt{ w } }  \right)
\label{pwjloclevy}
\end{eqnarray}
In particular it displays the power-law tail 
\begin{eqnarray}
{\cal P}^{loc}_{0<\mu<1} (w)  &&  \opsimeq_{w \to +\infty} \frac{\mu  \int dH_{jj} P_{diag}(H_{jj}) \vert H_{jj} - H_{j_0j_0}  \vert^{-\mu} }{2  N^{a \mu} w^{1+\frac{\mu}{2}}} 
\label{pwjloclevytail}
\end{eqnarray}
that translates for the exponent $
\alpha \equiv - \frac{ \ln w}{\ln N} $ into the multifractal spectrum 
for the probability
$\Pi^{loc} (\alpha) $ of $\alpha$
\begin{eqnarray}
\Pi^{loc}_{0<\mu<1} (\alpha)  &&  \simeq N^{  \frac{\alpha \mu }{2 } -a\mu  } 
\label{pchiloclevy}
\end{eqnarray}
The typical exponent corresponding to a finite probability $\Pi^{loc} (\alpha_{typ})  =O(1)$ is $\alpha_{typ}=2a$,
while the exponent associated to the maximal weight 
corresponding to a probability of order
 $\Pi^{loc} (\alpha_{min}) \sim \frac{1}{N}$
 is $\alpha_{min}=2 (a-\frac{1}{\mu})$.
So the number ${\cal N}^{loc}(\alpha)$ of weights scaling as $w \propto N^{-\alpha}$
\begin{eqnarray}
{\cal N}^{loc}_{0<\mu<1;a>\frac{1}{\mu}} (\alpha)  &&  \simeq N^{  \frac{\alpha \mu }{2 }  - (a \mu -1)  }   \theta \left( 2 (a-\frac{1}{\mu}) \leq  \alpha \leq 2a \right)
\label{dqloclevymusmall}
\end{eqnarray}

\subsection{ Critical point }

For the GRP model and (Eq. \ref{dqloc})
the L\'evy-GRP model for $1<\mu<2$ (Eq. \ref{dqloclevymubig}),
the critical point $a_c=1$ can be obtained as
 the limit $a \to a_c=1$ of the Localized phase,
and corresponds to the well known 'Strong Multifractality spectrum'
\cite{mirlin_fyodorov,mirlin_four}
\begin{eqnarray}
{\cal N}^{criti} (\alpha)  &&  \simeq N^{ \frac{\alpha}{2}  }   \theta \left( 0 \leq  \alpha \leq 2 \right)
\label{dqcriti}
\end{eqnarray}
 that appear in various Anderson Localization models
 (see the review \cite{mirlinrevue})
and that has been studied by various methods
 \cite{levitov1,levitov2,levitov3,levitov4,mirlin_evers,
fyodorov,fyodorovrigorous,oleg1,oleg2,oleg3,oleg4,oleg5,oleg6,olivier_per,olivier_strong,olivier_conjecture}.
It also appears in Many-Body-Localization models \cite{c_entropy,c_mblrgeigen}.

\subsection{ Discussion }

In summary, the perturbative expression 
of eigenvectors (Eq. \ref{eigenper}) 
is sufficient to derive the multifractal properties of the eigenstates
in the Localized phase and at criticality,
bur does not allow to go beyond the critical point $a_c=1$.
The goal of the present paper
 is thus to describe how the Wigner-Weisskopf approximation for the dynamics
yields a self-consistent perturbative expression for the eigenstates 
containing some broadening with respect to Eq \ref{eigenper},
in order to study the multifractal properties in the delocalized phase $a<a_c=1$.

\section{ Dynamics within the Wigner-Weisskopf approximation}

\label{sec_dyn}

\subsection{ Physical picture of the delocalized non-ergodic phase }

In this section, we describe the Wigner-Weisskopf approximation for the quantum dynamics from an initial site $j_0$ in order to obtain the inverse $\Gamma_{j_0}(N)$
 of the characteristic time to leave the state $j_0$
\begin{eqnarray}
\vert < j_0 \vert e^{-iHt} \vert j_0 > \vert^2 \simeq   e^{- \Gamma_{j_0} t }
\label{remainj0}
\end{eqnarray}
(here we have written the simplest exponential case, but we will also find the stretched exponential behavior in the L\'evy case).
The corresponding weights for the eigenvectors $\vert \phi_j > $ on the site $j_0$
\begin{eqnarray}
\vert < j_0 \vert \phi_j > \vert^2 \simeq   \frac{ \vert H_{j_0j} \vert^2 }{ (H_{jj}-H_{j_0j_0})^2 + \left( \frac{\Gamma_{j_0}(N)}{2} \right)^2 }  
\label{wj0}
\end{eqnarray}
then display the additional broadening $\Gamma_{j_0}(N) $
 with respect to Eq. \ref{wjloc}. 
When the broadening $\Gamma_{j_0}(N) $ is smaller in scaling than the level spacing $\Delta_{j_0}(N) \sim \vert H_{j_0j_0}-H_{next} \vert $,
one recovers the Localized phase with the weights of Eq. \ref{wjloc}. 
When the broadening $\Gamma_{j_0}(N) $ is bigger in scaling than the level spacing $\Delta_{j_0}(N) \sim \vert H_{j_0j_0}-H_{next} \vert $,
but remains smaller than the typical difference $\vert H_{j_0j_0}-H_{jj} \vert_{typ}  $, one obtains that the delocalization is only partial
since it involves the sub-extensive number $\frac{\Gamma_{j_0}(N)}{\Delta_{j_0}(N)} $ of states in the energy range $\Delta_{j_0}(N)  \leq \vert H_{j_0j_0}-H_{jj} \vert \leq   \Gamma_{j_0}(N) $.
The multifractal spectrum of eigenvectors can be then obtained from Eq. \ref{wj0}.
Note that here the broadening $ \Gamma_{j_0}(N)$ as determined by the dynamics (Eq. \ref{remainj0})
has a well-defined scaling in $N$ for each model as a function of its parameters. So this dynamical point of view is somewhat different from
 the closely recent studies based of the Green function $G(z)$ as a function of the complex variable $z=E+i \eta$,
where the imaginary part $\eta$ introduced as a formal regularization can be chosen with various scalings with respect to the system size $N$
in order to probe various regimes \cite{biroli_rosen,mirlin_tik,c_greenstat,ioffe}.

\subsection{ Dynamics from an initial site $j_0$     }

In terms of the components in the spatial basis
\begin{eqnarray}
\vert \psi(t)> = \sum_{j=1}^N \psi_j(t) \vert j >
\label{cjt}
\end{eqnarray}
the Schrodinger equation reads
\begin{eqnarray}
i \frac{d \psi_j(t) }{dt} = \sum_{k=1}^N H_{jk} \psi_k(t) = H_{jj} \psi_j(t)+\sum_{k \ne j} H_{jk} \psi_k(t)
\label{dynpsi}
\end{eqnarray}
with the initial condition 
\begin{eqnarray}
 \psi_j(t=0) = \delta_{jj_0}
\label{cj0}
\end{eqnarray}

It is convenient to work in the interaction picture, i.e. to make the change of variables
\begin{eqnarray}
\psi_j(t) = b_j(t) e^{-i H_{jj} t}
\label{cjbj}
\end{eqnarray}
so that Eq. \ref{dynpsi} becomes
\begin{eqnarray}
i \frac{d b_j(t) }{dt} =  \sum_{k \ne j} H_{jk} e^{i (H_{jj}-H_{kk})t } b_k(t)
\label{dynbj}
\end{eqnarray}
with the initial condition
\begin{eqnarray}
 b_j(t=0) = \delta_{jj_0}
\label{bj0}
\end{eqnarray}

\subsection{ First-order perturbation theory in the off-diagonal matrix elements  }

At order zero in the off-diagonal matrix elements, the solution is of course that the system remains forever in its initial condition $j_0$
\begin{eqnarray}
b^{(0)}_{j}(t)=\delta_{jj_0}
\label{ordrezero}
\end{eqnarray}
At first order,  the amplitudes on the other sites $j \ne j_0$ satisfy (Eq. \ref{dynbj}) 
\begin{eqnarray}
i \frac{d b_j^{(1)}(t) }{dt} =   H_{jj_0} e^{i (H_{jj}-H_{j_0j_0})t } 
\label{b1eq}
\end{eqnarray}
and thus read
\begin{eqnarray}
  b_j^{(1)}(t) && =   -i  H_{jj_0}  \int_0^t d\tau e^{i (H_{jj}-H_{j_0j_0}) \tau } 
%  \nonumber \\  && 
=i  H_{jj_0}  \frac{   2  \sin \left(  \frac{(H_{jj}-H_{j_0j_0})}{2} t\right)  }{ H_{j_0j_0}-H_{jj}}  e^{i \frac{(H_{jj}-H_{j_0j_0})}{2} t} 
\label{b1sol}
\end{eqnarray}
At lowest order, the probability to be still on the initial site $j_0$ at time $t$ will thus display the decay
\begin{eqnarray}
\vert b_{j_0}(t) \vert^2 \equiv 1- \gamma_{j_0}(t)
\label{b1squaresol}
\end{eqnarray}
where the probability to be elsewhere reads
\begin{eqnarray}
\gamma_{j_0} (t)&& = \sum_{j \ne j_0}  \vert b_j^{(1)}(t) \vert^2
=  \sum_{j \ne j_0}  
\left\vert  H_{jj_0}   \right\vert^2 f_t( H_{jj}-H_{j_0j_0} )
\label{gammasmalldef}
\end{eqnarray}
in terms of the well-known auxiliary function
\begin{eqnarray}
f_t (\omega ) = \left( \frac{     \sin \left(  \frac{\omega}{2} t\right)  }{ \frac{\omega}{2}   }  \right)^2
\label{ftomaga}
\end{eqnarray}
For large time $t$, this function becomes peaked around the origin
\begin{eqnarray}
f_t (\omega=0 ) = t^2
\label{ftomaga0}
\end{eqnarray}
on the interval $[- \frac{2 \pi}{t}, \frac{2 \pi}{t} ] $.

In the standard study of the decay into a continuum of states, this function is replaced by the delta function
\begin{eqnarray}
f_t (\omega ) \opsimeq_{t \to +\infty} 2 \pi t \delta(\omega)
\label{fregdelta}
\end{eqnarray}
and one obtains the famous Fermi Golden Rule 
\begin{eqnarray}
\gamma^{GR}_{j_0} (t) && \opsimeq_{t \to +\infty}   t \ \Gamma^{GR}_{j_0}
\label{fermi}
\end{eqnarray}
with the rate
\begin{eqnarray}
\Gamma^{GR}_{j_0} && = 2 \pi  \sum_{j=1}^N  \vert  H_{jj_0} \vert^2 \delta(H_{jj}-H_{j_0j_0})
\label{gammagolden}
\end{eqnarray}

Here since the states are discrete, one needs to keep the finite regularization of the delta function on the interval 
 $[- \frac{2 \pi}{t}, \frac{2 \pi}{t} ] $ leading to
\begin{eqnarray}
\gamma_{j_0} (t)&& \simeq t  2 \pi  \sum_{j=1}^N  \vert  H_{jj_0} \vert^2 \frac{  \theta\left(H_{j_0j_0}- \frac{2 \pi}{t}\leq  H_{jj} \leq  H_{j_0j_0}+ \frac{2 \pi}{t} \right)  }{ \frac{4 \pi}{t} }
\nonumber \\
&&=\frac{t^2}{2}    \sum_{j=1}^N  \vert  H_{jj_0} \vert^2  \theta\left(H_{j_0j_0}- \frac{2 \pi}{t}\leq  H_{jj} \leq  H_{j_0j_0}+ \frac{2 \pi}{t} \right) 
\label{gammasmall}
\end{eqnarray}

It is thus useful to introduce the number of resonances the interval 
 $[- \frac{2 \pi}{t}, \frac{2 \pi}{t} ] $
\begin{eqnarray}
N_t \equiv  \sum_{j=1}^N   \theta\left(H_{j_0j_0}- \frac{2 \pi}{t}\leq  H_{jj} \leq  H_{j_0j_0}+ \frac{2 \pi}{t} \right) 
\label{nt}
\end{eqnarray}
As long as this number remains large $N_t \gg 1$, it will concentrate around its averaged value
\begin{eqnarray}
N_t \simeq  N   \int_{H_{j_0j_0}- \frac{2 \pi}{t}}^{H_{j_0j_0}+ \frac{2 \pi}{t} } dH_{jj} P_{diag}(H_{jj})
\opsimeq_{t \gg 1}   N    P_{diag}(H_{j_0j_0})  \frac{4 \pi}{t}
\label{ntbis}
\end{eqnarray}
involving the probability density $P_{diag}(H_{j_0j_0}) $ of the diagonal element $ H_{j_0j_0}$.

\subsection{ Wigner-Weisskopf approximation  }

As explained in quantum mechanics textbooks,
 the Wigner-Weisskopf approximation 
allows to promote the linear perturbative decay of Eq. \ref{b1squaresol} 
into an exponential decay as follows.

Eq \ref{dynbj} is written exactly for the initial site $j_0$
\begin{eqnarray}
i \frac{d b_{j_0}(t) }{dt} =  \sum_{j \ne j_0} H_{j_0j} e^{i (H_{j_0j_0}-H_{jj})t } b_j(t)
\label{dynnini}
\end{eqnarray}
while for the other sites $j \ne j_0$, one keeps only the dominant term $k=j_0$ on the right hand-side
\begin{eqnarray}
i \frac{d b_j(t) }{dt} =  H_{jj_0} e^{i (H_{jj}-H_{j_0j_0})t } b_{j_0}(t)
\label{dynnother}
\end{eqnarray}
The integration
\begin{eqnarray}
  b_j(t) &&  =  -i  H_{jj_0}  \int_0^t d\tau e^{i (H_{jj}-H_{j_0j_0}) \tau } b_{j_0}(\tau)
\label{bjt}
\end{eqnarray}
is plugged into Eq. \ref{dynnini} to obtain a closed equation for the amplitude at $j_0$
\begin{eqnarray}
i \frac{d b_{j_0}(t) }{dt} && \simeq -i \sum_{j \ne j_0} \vert H_{j_0j} \vert^2  
   \int_0^t d\tau e^{ i (H_{j_0j_0}-H_{jj}) (t-\tau) } b_{j_0}(\tau)
\nonumber \\
&& \simeq -i \sum_{j \ne j_0} \vert H_{j_0j} \vert^2   
   \int_0^t dx e^{ i (H_{j_0j_0}-H_{jj}) x } b_{j_0}(t-x)
\label{closed}
\end{eqnarray}
To simplify further, one makes the Markovian approximation $b_{0}(t-x) \simeq b_{0}(t) $
to obtain
\begin{eqnarray}
\frac{1}{b_{j_0}(t)} \frac{d b_{j_0}(t) }{dt} 
&& \simeq  - \sum_{j \ne j_0} \vert H_{j_0j} \vert^2   
   \int_0^t dx e^{ i (H_{j_0j_0}-H_{jj}) x } 
\label{markov}
\end{eqnarray}

The integration with the initial condition $b_{j_0}(t=0)=1$ yields
\begin{eqnarray}
\ln b_{j_0}(t) 
 && \simeq  - \sum_{j \ne j_0} \vert H_{j_0j} \vert^2   
  \int_0^t d\tau  \int_0^{\tau} dx e^{ i (H_{j_0j_0}-H_{jj}) x } 
\nonumber \\
&& \simeq   - \sum_{j \ne j_0} \vert H_{j_0j} \vert^2   \frac{1+i (H_{j_0j_0}-H_{jj}) t -e^{i (H_{j_0j_0}-H_{jj}) t} }{(H_{j_0j_0}-H_{jj})^2}
\nonumber \\
&& \simeq   - \frac{1}{2} \sum_{j \ne j_0} \vert H_{j_0j} \vert^2   \frac{ \sin^2 \left(  \frac{(H_{jj}-H_{j_0j_0})}{2} t\right)  }{  \left(  \frac{(H_{jj}-H_{j_0j_0})}{2} \right) ^2}
 - i \sum_{j \ne j_0} \vert H_{j_0j} \vert^2   \frac{ (H_{j_0j_0}-H_{jj}) t - \sin [ (H_{j_0j_0}-H_{jj}) t] }{(H_{j_0j_0}-H_{jj})^2}
\label{dyn0}
\end{eqnarray}
The first real term involves the function $\gamma_{j_0} (t) $ already introduced in Eq \ref{gammasmalldef},
while the second imaginary term is dominated by the contribution which is linear in time,
where the coefficient
\begin{eqnarray}
 \delta_{j_0}
 && = \sum_{j \ne j_0} \frac{ \vert H_{j_0j} \vert^2   }{H_{j_0j_0}-H_{jj}}
\label{deltaper}
\end{eqnarray}
is well-known as the second-order perturbation correction to the eigenvalue $H_{j_0j_0}$.

In summary, the amplitude on the initial site $j_0$ of Eq. \ref{dyn0}
follows the exponential form
\begin{eqnarray}
   b_{j_0}(t) \simeq e^{- \frac{\gamma_{j_0} (t) }{2} -i \delta_{j_0} t }
\label{solubj0}
\end{eqnarray}
while the amplitudes on the other sites $j \ne j_0$ become (Eq. \ref{bjt})
\begin{eqnarray}
  b_j(t) &&  \simeq  -i  H_{jj_0}  \int_0^t d\tau  e^{- \frac{\gamma_{j_0 } (\tau) }{2} - i ( H_{j_0j_0}-H_{jj} +  \delta_{j_0} ) \tau   }
\label{bjtsolu}
\end{eqnarray}
In particular, this Wigner-Weisskopf approximation yields the final probabilities of the other states $j \ne j_0$
in the limit $t \to +\infty$
\begin{eqnarray}
\vert  \psi_j(t \to +\infty) \vert^2 =\vert  b_j(t \to +\infty) \vert^2 &&  \simeq  \left\vert  -i  H_{jj_0}  \int_0^{+\infty} d\tau  e^{- \frac{\gamma_{j_0 } (\tau) }{2} - i ( H_{j_0j_0}-H_{jj} +  \delta_{j_0} ) \tau   } \right\vert^2
\label{bjasympproba}
\end{eqnarray}

\subsection{ Interpretation from the point of view of the eigenstates  }

For the amplitudes $\psi_j(t)$ of Eq \ref{cjt}, 
the solution of Eq. \ref{bjtsolu} yields via Eq \ref{cjbj}
\begin{eqnarray}
  \psi_j(t) &&   \opsimeq_{t \to +\infty}     e^{-i H_{jj} t } b_j(\infty)
\label{psijtsolu}
\end{eqnarray}
The comparison with the spectral decomposition into eigenstates
\begin{eqnarray}
  \vert\psi(t)> = \sum_{n=1}^N    e^{-i E_n t } \vert \phi_n >< \phi_n \vert j_0 >
\label{psiteigen}
\end{eqnarray}
means that at this approximation, the eigenvalues are $E_{_j}=H_{jj}+...$, the corresponding eigenstates are 
 $\vert \phi_{j}> =\vert j>+...$, so that the amplitudes of these eigenstates at $j_0$ can be identified to
\begin{eqnarray}
< \phi_{j} \vert j_0 > =  b_j(\infty) \simeq  -i  H_{jj_0}  \int_0^{+\infty} d\tau  e^{- \frac{\gamma_{j_0 } (\tau) }{2} - i ( H_{j_0j_0}-H_{jj} +  \delta_{j_0} ) \tau   }
\label{amplinj0}
\end{eqnarray}

\subsection{  Example with the exponential decay $\gamma_{j_0}(t) = \Gamma_{j_0} t  $  }

\label{sec_linearexp}

The exponential decay $\gamma_{j_0}(t) = \Gamma_{j_0} t  $
corresponds to the standard Golden-Rule form (Eq. \ref{fermi})
and to the standard Wigner-Weisskopf approximation,
where the amplitudes of Eq. \ref{amplinj0} 
\begin{eqnarray}
 < \phi_{j} \vert j_0 >   
&& \simeq    \frac{  H_{jj_0}    }
{( H_{jj} - H_{j_0j_0} - \delta_{j_0} )  +   i \frac{\Gamma_{j_0 } }{2} }
\label{ampli}
\end{eqnarray}
lead to the well-known Lorentzian shape for the weights
\begin{eqnarray}
\vert   < \phi_{j} \vert j_0 >\vert^2 &&  \simeq  \frac{  \vert  H_{jj_0}  \vert^2}
{( H_{jj} - H_{j_0j_0} - \delta_{j_0} )^2  +   \left( \frac{\Gamma_{j_0 } }{2} \right)^2 }
\label{lorentz}
\end{eqnarray}

\subsection{  Example with the stretched exponential decay $\gamma_{j_0}(t) = (\Gamma_{j_0} t )^{\beta} $  with $0<\beta<1$ }

\label{sec_streched}

For the stretched exponential decay $\gamma_{j_0}(t) = (\Gamma_{j_0} t )^{\beta} $  with $0<\beta<1$, the weights
\begin{eqnarray}
\vert  < \phi_{j} \vert j_0 >  \vert^2 &&  \simeq  \left\vert  H_{jj_0}  \int_0^{+\infty} d\tau  e^{-  \frac{(\Gamma_{j_0 } \tau)^{\beta} }{2} -i ( H_{j_0j_0}  -H_{jj}+  \delta_{j_0} ) \tau } \right\vert^2 =   \left\vert  H_{jj_0}  I_{\beta}( H_{j_0j_0}  -H_{jj}+  \delta_{j_0};\Gamma_{j_0})\right\vert^2
\label{wstreched}
\end{eqnarray}
involve the half-Fourier of a stretched exponential
\begin{eqnarray}
I_{\beta}(\omega;\Gamma) \equiv   \int_0^{+\infty} d\tau  e^{-  \frac{(\Gamma \tau)^{\beta} }{2} -i \omega \tau } 
\label{integrale}
\end{eqnarray}
which does not seem to have a simple explicit expression (while the full Fourier corresponds to the L\'evy symmetric stable law
of index $\beta$).
However the stretched exponential can be rewritten as the Laplace transform of the fully asymmetric L\'evy stable law $L_{\beta}(x)$
of index $\beta$
\begin{eqnarray}
  e^{-  \frac{(\Gamma \tau)^{\beta} }{2}  } = \int_0^{+\infty} dx L_{\beta}(x)  e^{-  (\Gamma 2^{-\frac{1}{\beta}}) \tau x }
\label{levyasym}
\end{eqnarray}
so that Eq. \ref{integrale} become
\begin{eqnarray}
I_{\beta}(\omega;\Gamma) && = \int_0^{+\infty} dx L_{\beta}(x)   \int_0^{+\infty} d\tau
  e^{ - i \omega \tau }   e^{-  (\Gamma 2^{-\frac{1}{\alpha}}) \tau x }
%  \nonumber \\  && 
= \int_0^{+\infty} dx L_{\beta}(x)   \frac{1}{ (\Gamma 2^{-\frac{1}{\beta}})  x  + i \omega}  
\nonumber \\
&& = \int_0^{+\infty} dx L_{\beta}(x)   \frac{  (\Gamma 2^{-\frac{1}{\beta}})  x  }{ (\Gamma 2^{-\frac{1}{\beta}})^2  x^2  +  \omega^2}  
-i \int_0^{+\infty} dx L_{\beta}(x)   \frac{    \omega}{ (\Gamma 2^{-\frac{1}{\beta}})^2  x^2  +  \omega^2}  
\label{integrale2}
\end{eqnarray}
and one obtains the weights of Eq. \ref{wstreched} in terms of these integrals.

However, in the following we will only need the two simple limits :

(i) for $\Gamma \ll \vert \omega \vert $, we may approximate by the value for $\Gamma_{j_0}\to 0$
\begin{eqnarray}
\vert I_{\beta}(\omega;\Gamma \to 0) \vert^2 =  \frac{1}{\omega^2} 
\label{integralegammavanish}
\end{eqnarray}
so that the weights of Eq. \ref{wstreched} become
\begin{eqnarray}
\vert  < \phi_{j} \vert j_0 >  \vert^2 && \ \ \    \opsimeq_{\Gamma_{j_0} \ll  \vert H_{j_0j_0}  -H_{jj}\vert }  
\ \ \ \  \left\vert \frac{ H_{jj_0}  }
{ H_{j_0j_0}  -H_{jj} } \right\vert^2
\label{wstrechedloc}
\end{eqnarray}
as it should to recover Eq. \ref{eigenper}.

(ii) for $\Gamma \gg \vert \omega \vert $, we may approximate by the value for $\omega=0$
\begin{eqnarray}
\vert I_{\beta}(\omega=0;\Gamma ) \vert^2 =  \frac{ \left[  2^{\frac{1}{\beta}}  \int_0^{+\infty} du u^{\frac{1}{\beta}} e^{-u}  \right]^2 }{\Gamma^2} 
\label{integraleomegavanish}
\end{eqnarray}
so that the weights of Eq. \ref{wstreched} reads
\begin{eqnarray}
\vert  < \phi_{j} \vert j_0 >  \vert^2 && \ \ \    \opsimeq_{\Gamma_{j_0} \gg  \vert H_{j_0j_0}  -H_{jj}\vert }  
\ \ \ \  \left[  2^{\frac{1}{\beta}}  \int_0^{+\infty} du u^{\frac{1}{\beta}} e^{-u}  \right]^2   \left\vert \frac{ H_{jj_0}  }
{ \Gamma_{j_0}  } \right\vert^2
\label{wstrechedinside}
\end{eqnarray}
i.e. apart from numerical constants, the energy difference $\vert  H_{j_0j_0}  -H_{jj} \vert $ of Eq. \ref{wstrechedloc}
is simply replaced by the broadening $ \Gamma_{j_0}$, exactly as in the Lorentzian simpler case of Eq. \ref{lorentz}.

\section{ Generalized-Rosenzweig-Porter matrix model }

\label{sec_rosen}

As recalled in the Introduction, the Generalized-Rosenzweig-Porter model is the simplest matrix model exhibiting a delocalized non-ergodic phase
with an explicit multifractal spectrum for eigenvectors in \cite{kravtsov_rosen}, and has been analyzed
recently from various points of view  \cite{biroli_rosen,ossipov_rosen,ioffe}.
In this section, our goal is to show how the present dynamical approach is able to recover the multifractal spectrum obtained in \cite{kravtsov_rosen}.

\subsection{ Dynamics within the Wigner-Weisskopf approximation }

Here the number of resonances of Eq. \ref{nt} scales as Eq \ref{ntbis}
\begin{eqnarray}
N_t \simeq     N    P_{diag}(H_{j_0j_0})  \frac{4 \pi}{t}
\label{ntbisk}
\end{eqnarray}
Since all off-diagonal matrix elements have the same scaling (Eq. \ref{offa} and \ref{rosenvar}), Eq. \ref{gammasmall} becomes
\begin{eqnarray}
\gamma_{j_0} (t)&& \simeq \frac{t^2}{2 N^{2a} }   N_t  \simeq    2 \pi  P_{diag}(H_{j_0j_0})      N^{1-2a}  t
\label{gammasmallrosen}
\end{eqnarray}
It is thus linear in the time $t$ as the case discussed in section \ref{sec_linearexp}, 
leading to the Lorentzian weights (Eq \ref{lorentz})
\begin{eqnarray}
w_j \equiv \vert   < \phi_{j} \vert j_0 >\vert^2 &&  \simeq  \frac{  \vert  H_{jj_0}  \vert^2}
{( H_{jj} - H_{j_0j_0} - \delta_{j_0} )^2  +   \left( \frac{\Gamma_{j_0 } }{2} \right)^2 }
\label{lorentzk}
\end{eqnarray}
with the broadening 
\begin{eqnarray}
\Gamma_{j_0}(N) && = 2 \pi   P_{diag}(H_{j_0j_0})     N^{1-2 a}   
\label{Gammarosen}
\end{eqnarray}
that should be compared with the level spacing $\Delta_{j_0}(N)$ of
Eq \ref{levelspacing}.
For $a>a_c=1$, the broadening $\Gamma_{j_0}(N) $ is smaller in scaling
than the level spacing $\Delta_{j_0}(N)$ and one recovers the localized
phase discussed in section \ref{sec_rosenloc}.

\subsection{ Multifractality in the delocalized non-ergodic phase $\frac{1}{2}<a<a_c=1$ }

For $\frac{1}{2}<a<a_c=1$, the broadening $\Gamma_{j_0}(N) $ of Eq. \ref{Gammarosen} decays with $N$ but
 is bigger in scaling than the level spacing $\Delta_{j_0} (N) $ of Eq. \ref{levelspacing}, so here we need to analyze the Lorentzian weights
\begin{eqnarray}
w_j  &&  \simeq  \frac{  \vert  H_{jj_0}  \vert^2}
{( H_{jj} - H_{j_0j_0}  )^2  +   \left( \frac{\Gamma_{j_0 } }{2} \right)^2 } 
=  \frac{  N^{-2a }  }
{( H_{jj} - H_{j_0j_0}  )^2  +   \left( \frac{\Gamma_{j_0 } }{2} \right)^2 }
\label{lorentzkk}
\end{eqnarray}
The typical value remains the same as in Eq. \ref{wjloctyp},
\begin{eqnarray}
w_j^{typ} &&  \propto  N^{-2a} 
\label{wjloctypbis}
\end{eqnarray}
while the maximal weight is not Eq. \ref{wjlocmax} anymore but is given instead by 
\begin{eqnarray}
w^{max}(N) &&  \equiv 4 \frac{  N^{-2a }  }{\Gamma_{j_0 } ^2 } \propto   N^{ -2 (1-a) }
\label{wjlmax}
\end{eqnarray}

In terms of this maximal value $ w^{max}(N)$ , the probability distribution reads
\begin{eqnarray}
{\cal P} (w)  &&  = \int dH_{jj} P_{diag}(H_{jj}) \delta \left( w-  \frac{  N^{-2a} } {( H_{jj} - H_{j_0j_0} )^2+   \left( \frac{\Gamma_{j_0 } }{2} \right)^2  } \right)
\nonumber \\
&& = \theta(w \leq w^{max}(N) ) \frac{P_{diag}\left(H_{j_0j_0} + N^{-a} \sqrt{\frac{1}{ w } -\frac{1}{ w^{max}(N) } } \right) 
+ P_{diag}\left(H_{j_0j_0} - N^{-a} \sqrt{\frac{1}{ w } -\frac{1}{ w^{max}(N) } } \right)  }
{ 2 N^a w^{\frac{3}{2}}  \sqrt{ 1- \frac{w}{w^{max}(N)}  } }
\label{pwjnon}
\end{eqnarray}

For the exponent $\alpha=-\frac{\ln w}{\ln N}$, this translates into the multifractal spectrum for the number ${\cal N}(\alpha)$ 
\begin{eqnarray}
{\cal N}^{nonergo}_{\frac{1}{2}<a<1}(\alpha) &&  \simeq  N^{ \frac{\alpha}{2} +1-a }  \  \theta(   2(1-a) \leq \alpha  \leq  2a )  
\label{pchi}
\end{eqnarray}
The physical meaning of this delocalized non-ergodic phase is thus as follows : 
the delocalization is limited to the energies inside the broadening scale $\vert H_{jj} - H_{j_0j_0} \vert < \Gamma_{j_0}(N) \propto N^{1-2 a}   $ 
containing the sub-extensive $\frac{\Gamma_{j_0}(N)}{\Delta_{j_0}(N)} \propto N^{2(1- a)}$ number of states
that have weights scaling as $w^{max}(N)  \propto   N^{ -2 (1-a) } $ (Eq. \ref{wjlmax}).
The other exponents $\alpha> 2(1-a) $ arising in the linear spectrum of Eq. \ref{pchi}
corresponds to energies outside the broadening scale $\vert H_{jj} - H_{j_0j_0} \vert < \Gamma_{j_0}(N) \propto N^{1-2 a}   $.
For the generalized fractal dimensions $D(q)$ that govern the 
generalized moments of arbitrary index $q>0$
\begin{eqnarray}
N <w^q>_N = \int d \alpha {\cal N}(\alpha)  N^{- \alpha q}  \simeq \ \  N^{(1-q) D(q) }
\label{defdq}
\end{eqnarray}
Eq. \ref{pchi} translates into
\begin{eqnarray}
D^{nonergo}_{\frac{1}{2}<a<1}(q) && =  2(1-a)   \ \ \  \ \ \ \  {\rm for } \ \ \ q \geq \frac{1}{2}
\nonumber \\
D^{nonergo}_{\frac{1}{2}<a<1}(q) && =  \frac{1- 2aq}{1-q}  \ \ \  \ \ \ {\rm for } \ \ \ 0 \leq q \leq \frac{1}{2}
\label{dqnonergo}
\end{eqnarray}
So the region inside the broadening scale $\vert H_{jj} - H_{j_0j_0} \vert < \Gamma_{j_0}(N)   $ govern all the fractal dimensions $D(q)$ for $q>\frac{1}{2}$,
while the region outside the broadening scale $\vert H_{jj} - H_{j_0j_0} \vert < \Gamma_{j_0}(N)   $ dominates for $q<\frac{1}{2}$.

It is interesting to consider the two boundaries of the delocalized non-ergodic region $\frac{1}{2}<a<a_c=1$.
For $a \to a_c=1$, one recovers the critical spectrum of Eq. \ref{dqcriti} as it should.
For $a \to \frac{1}{2}$, one reaches the monofractal spectrum of the ergodic phase
\begin{eqnarray}
{\cal N}^{ergo}_{a=\frac{1}{2}}(\alpha) &&  \simeq  N  \delta( \alpha -1 )  
\label{multifergo}
\end{eqnarray}
Note that for this case $a=\frac{1}{2}$ where the broadening $\Gamma_{j_0}$ does not decay with $N$ anymore,
 the Lorentzian distribution of Eq. \ref{lorentzkk}
is nevertheless a non-perturbative exact result
as a consequence of the free probability theory as applied to eigenvectors (see \cite{free_add,ithier} and references therein).

\section{  L\'evy version of the Generalized-Rosenzweig-Porter matrix model }

\label{sec_levy}

\subsection{ Dynamics within the Wigner-Weisskopf approximation
  in the region $1<\mu<2$  }

The sum of Eq. \ref{gammasmall} that we have to evaluate
\begin{eqnarray}
\gamma_{j_0} (t)
&&=\frac{t^2}{2}    \sum_{j=1}^N  \vert  H_{jj_0} \vert^2  \theta\left(H_{j_0j_0}- \frac{2 \pi}{t}\leq  H_{jj} \leq  H_{j_0j_0}+ \frac{2 \pi}{t} \right) 
\label{gammasmalllevy}
\end{eqnarray}
involves the number
(Eq \ref{nt} and \ref{ntbis}) 
\begin{eqnarray}
N_t \equiv  \sum_{j=1}^N   \theta\left(H_{j_0j_0}- \frac{2 \pi}{t}\leq  H_{jj} \leq  H_{j_0j_0}+ \frac{2 \pi}{t} \right) 
\simeq    N    P_{diag}(H_{j_0j_0})  \frac{4 \pi}{t}
\label{ntlevy0}
\end{eqnarray}
of random positive variables $y_j \equiv \vert  H_{jj_0} \vert^2 $, 
whose distribution is obtained from Eq. \ref{tailoff} 
\begin{eqnarray}
 {\cal P}(y_j) =
  \frac{ \mu }{ 2  N^{a\mu} y_j^{1+\frac{\mu}{2}} } 
\theta \left( y_j \geq N^{-2a} \right)
\label{tailyj}
\end{eqnarray}
As a consequence, the sum $S_{N_t}$ of $N_t$ variables $y_j$ 
is distributed with the asymmetric L\'evy stable distribution
of index $\frac{\mu}{2}$. In particular displays the tail
\begin{eqnarray}
 {\cal P}(S_{N_t}) \opsimeq_{S_{N_t} \to +\infty}
  \frac{ \mu N_t }{ 2  N^{a\mu} S_{N_t}^{1+\frac{\mu}{2}} } 
\label{tailst}
\end{eqnarray}
so that its typical scaling reads
\begin{eqnarray}
 S^{typ}_{N_t}  \simeq
\left(  \frac{  \mu  N_t   }{ 2 N^{a\mu}   }  \right)^{\frac{2}{\mu}} 
\label{stypnt}
\end{eqnarray}

Putting everything together,  Eq \ref{gammasmalllevy} scales as
\begin{eqnarray}
\gamma_{j_0} (t)  \simeq \frac{t^2}{2}  S^{typ}_{N_t}
%% \nonumber \\ && 
 \simeq \frac{t^{2-\frac{2}{\mu}} }{2}  \left(  \frac{  \mu       P_{diag}(H_{j_0j_0}) 2 \pi   }{  N^{a\mu-1}   }  \right)^{\frac{2}{\mu}}
  \equiv \frac{ (\Gamma_{j_0}t)^{\beta} }{2}
\label{gammasmalllevyres}
\end{eqnarray}
This corresponds to the stretched exponential case discussed in section \ref{sec_streched}
with the exponent  
\begin{eqnarray}
\beta = \frac{2}{\mu} (\mu-1) 
\label{alphalevy}
\end{eqnarray}
varying in the interval $0<\beta<1$ for $1<\mu<2$.
The inverse time scale in Eq. \ref{gammasmalllevyres}
\begin{eqnarray}
\Gamma_{j_0} =
  \frac{ \left[  \mu P_{diag}(H_{j_0j_0}) 2 \pi  \right]^{\frac{1}{\mu-1}}   }
{  N^{\frac{a\mu-1}{\mu-1}}   }  
\label{gammalevyres}
\end{eqnarray}
decays as a function of the system size $N$ for $a>\frac{1}{\mu}$.
The comparison with the level spacing 
$\Delta_{j_0} (N) \propto \frac{1}{N \rho(H_{j_0j_0})}$
shows that the delocalized non-ergodic phase corresponds to the region
\begin{eqnarray}
\frac{1}{\mu}<a^{nonergo}<a_c=1
\label{nonergolevy}
\end{eqnarray}

\subsection{ Multifractal properties in the delocalized non-ergodic
region $\frac{1}{\mu}<a<a_c=1 $ for $1<\mu<2$  }

In the region of Eq. \ref{nonergolevy},
the broadening $\Gamma_{j_0}(N)$ of Eq. \ref{gammalevyres} decays with $N$ but is bigger in scaling than the level 
spacing of Eq. \ref{levelspacing}.
As explained in section \ref{sec_streched}, the weights of the eigenstates are more complicated than Lorentzian,
but to obtain the multifractal spectrum, we only need to take into account the two simple limits
of Eq. \ref{wstrechedloc} and Eq. \ref{wstrechedinside} as follows :

(i)   In the region outside the broadening scale  $ \vert H_{jj} - H_{j_0j_0} \vert >  \Gamma_{j_0 } $,
the weights still follow Eq. \ref{wstrechedloc}
\begin{eqnarray}
w_j^{outside} \simeq  \frac{ H_{jj_0}^2  }{ (H_{j_0j_0}  -H_{jj} )^2 } = \frac{ N^{-2a} v^2  }{ (H_{j_0j_0}  -H_{jj} )^2 }
\label{wstrechedlocout}
\end{eqnarray}

Using Eq. \ref{pofflevy},
the probability distribution of these weights reads
\begin{eqnarray}
{\cal P}^{outside} (w)  &&  = \int dH_{jj} P_{diag}(H_{jj}) \int dv 
 \frac{ \mu }{ 2  \vert v \vert^{1+\mu} }  \theta \left( \vert v \vert \geq 1 \right)
 \theta(\vert H_{jj} - H_{j_0j_0} \vert >  \Gamma_{j_0 }  )   \delta \left( w-  \frac{  N^{-2a} v^2 } {( H_{jj} - H_{j_0j_0} )^2  } \right)
\nonumber \\
&& = \frac{1}{2 N^a w^{\frac{3}{2} }}  \int dv 
 \frac{ \mu }{ 2  \vert v \vert^{\mu} }  \theta \left( \vert v \vert \geq 1 \right)
 \theta(    w \leq \frac{ N^{-2a} v^2  }{ \Gamma_{j_0 } ^2 } )
% \nonumber \\  &&  
  \left[   P_{diag} \left(  H_{j_0j_0}- v \frac{  N^{-a}  } { \sqrt{w} } \right) +P_{diag} \left(  H_{j_0j_0}+ v\frac{  N^{-a}  } { \sqrt{w} } \right) \right]
\nonumber 
\label{pwjnonergolevyoutside}
\end{eqnarray}

This translates into the multifractal spectrum for probability distribution of
the exponent $\alpha=-\frac{\ln w}{\ln N}$
\begin{eqnarray}
\Pi^{outside}(\alpha) \simeq  N^{ \frac{\alpha}{2} -a  } \theta \left(  2  \frac{(1-a)}{(\mu-1)} \leq \alpha \leq  2a \right)
\label{pialphaoutside}
\end{eqnarray}
Since there are $O(N)$ weights outside, the corresponding number of weights decaying with the exponent $\alpha$ reads
\begin{eqnarray}
{\cal N}^{outside}(\alpha) \simeq  N^{ \frac{\alpha}{2} +1-a  } \theta \left(  2  \frac{(1-a)}{(\mu-1)} \leq \alpha \leq  2a \right)
\label{nalphaoutside}
\end{eqnarray}

ii) In the region inside the broadening scale $ \vert H_{jj} - H_{j_0j_0} \vert <  \Gamma_{j_0 }  $, the number of states scales as
\begin{eqnarray}
 \frac{\Gamma_{j_0 }}{\Delta_{j_0} } \propto  N^{1- \frac{a\mu-1}{\mu-1}} = N^{\mu \frac{1-a}{\mu-1}}
\label{insidelevy}
\end{eqnarray}
and the weights follow Eq. \ref{wstrechedinside}
\begin{eqnarray}
w_j^{inside} && \simeq  \frac{ H_{jj_0}^2 } { \Gamma_{j_0}^2 } =  \frac{ N^{-2a} v^2 } { \Gamma_{j_0}^2 } 
 \propto N^{-2  \frac{1-a}{\mu-1}}  v^2
\label{wstrechedinsidelevy}
\end{eqnarray}

Using Eq. \ref{pofflevy},
the probability distribution of these weights reads
\begin{eqnarray}
{\cal P}^{inside} (w)  &&  =  \int dv 
 \frac{ \mu }{ 2  \vert v \vert^{1+\mu} }  \theta \left( \vert v \vert \geq 1 \right)   \delta \left( w-  \frac{ N^{-2a} v^2 } { \Gamma_{j_0}^2 }   \right)
\nonumber \\
&& =\frac{ \mu }{ 2 w^{1+\frac{\mu}{2}} } N^{-a \mu} \Gamma_{j_0}^{-\mu}   \theta  \left( w \geq   \frac{ N^{-2a}  } { \Gamma_{j_0}^2 }   \right)
\label{pwjnonergolevyinside}
\end{eqnarray}

Using that $ \Gamma_{j_0 } $ decays with $N$ as Eq. \ref{gammalevyres},
this translates into the multifractal spectrum for probability distribution of
the exponent $\alpha=-\frac{\ln w}{\ln N}$
\begin{eqnarray}
\Pi^{inside}(\alpha) \simeq  N^{\alpha \frac{\mu}{2} - \mu \frac{1-a}{\mu-1} } \theta \left(   \alpha \leq 2  \frac{1-a}{\mu-1} \right)
\label{pialphainside}
\end{eqnarray}
Since the number of states inside the broadening scales as Eq. \ref{insidelevy}, the corresponding number of exponents $\alpha$ 
\begin{eqnarray}
{\cal N}^{inside}(\alpha) \simeq  N^{\alpha \frac{\mu}{2}  }  \theta \left( 0  \leq  \alpha \leq 2  \frac{(1-a)}{(\mu-1)} \right)
\label{nalphainside}
\end{eqnarray}

Putting together the two contributions of Eq. \ref{pialphaoutside}
and Eq \ref{pialphainside}, one obtains that the total multifractal spectrum is the sum of two linear spectra of slopes $\frac{\mu}{2}$ and $\frac{1}{2}$
\begin{eqnarray}
{\cal N}^{nonergo}_{1<\mu<2;\frac{1}{\mu}<a<1}(\alpha) && ={\cal N}^{inside}(\alpha)  + {\cal N}^{outside}(\alpha) 
\nonumber \\
&& =  N^{\alpha \frac{\mu}{2}  }  \theta \left( 0  \leq  \alpha \leq 2  \frac{(1-a)}{(\mu-1)} \right)
+  N^{ \frac{\alpha}{2} +1-a  } \theta \left(  2  \frac{(1-a)}{(\mu-1)} \leq \alpha \leq  2a \right)
\label{pialphatotal}
\end{eqnarray}
For the generalized fractal dimensions $D(q)$ that govern the 
generalized moments of arbitrary index $q>0$ (Eq. \ref{defdq}),
Eq. \ref{pialphatotal} translates into the three domains
\begin{eqnarray}
D^{nonergo}_{1<\mu<2;\frac{1}{\mu}<a<1}(q) && =  0  \ \ \ \ \ \ \  \  \ \ \  \ \ \ \  \ \ \ \ \  \  \ \ \ \ \ \ \ \ \  \  \ \ \ {\rm for } \ \  \ 
q \geq \frac{\mu}{2}
\nonumber \\
D^{nonergo}_{1<\mu<2;\frac{1}{\mu}<a<1}(q) && =   \frac{ \mu \frac{(1-a)}{(\mu-1)} \left(1- \frac{2}{\mu} q \right)  }{1-q}  \ \ \ \ \ \  \  \ \ \  \ \ \ \ \ {\rm for } \ \ \ 
\frac{1 }{2} \leq q \leq \frac{\mu}{2}
\nonumber \\
D^{nonergo}_{1<\mu<2;\frac{1}{\mu}<a<1}(q)  && =  \frac{1- 2a q }{1-q}  \ \ \ \ \ \ \ \ \ \ \ \ \ \ \  \ \ \ \  \ \ \ \  {\rm for } \ \ \ 0 \leq q \leq \frac{1 }{2} 
\label{multiflevy}
\end{eqnarray}
The results are thus very different
 from the spectrum of Eq. \ref{dqnonergo} concerning the Generalized-Rosenzweig-Potter.
The 'delocalization' for the energies inside the broadening scale $\vert H_{jj} - H_{j_0j_0} \vert < \Gamma_{j_0}(N)  $ 
containing the sub-extensive $\frac{\Gamma_{j_0}(N)}{\Delta_{j_0}(N)} $ number of states 
is not homogeneous as in the  Generalized-Rosenzweig-Potter, 
but is instead strongly inhomogeneous as
a consequence of the L\'evy distribution of the off-diagonal matrix elements. So this 'delocalization' is actually not so effective.
In particular the generalized dimensions $D(q)$ vanish in the whole region $q>\frac{\mu}{2} $ including the information dimension $D(q=1)$
and the return dimension $D(q=2)$, while the transmission dimension does not vanish $D(q=\frac{1}{2})>0$. 
Our conclusion is thus that the possibility proposed by Cizeau and Bouchaud \cite{cizeau} 
to have at the same time $D(q=\frac{1}{2})>0$ ($\Upsilon=\infty $ in the notation of \cite{cizeau}) and $D(q=2)=0$ 
($ Y>0$ in the notation of \cite{cizeau}) indeed comes true for the present L\'evy-GRP model.

At the critical point $a_c=1$, Eq. \ref{pialphatotal} yields the critical spectrum of Eq. \ref{dqcriti} corresponding to the contribution $ {\cal N}^{outside}(\alpha)  $ only
\begin{eqnarray}
{\cal N}^{criti}_{1<\mu<2;a_c=1}(\alpha) &&  = N^{ \frac{\alpha}{2}   } \theta \left(  0 \leq \alpha \leq  2a \right)
\label{pialphatotalcriti}
\end{eqnarray}
At the other boundary $a=\frac{1}{\mu} $ on the contrary, only the contribution ${\cal N}^{inside}(\alpha) $ survives and gives
\begin{eqnarray}
{\cal N}^{nonergo}_{1<\mu<2;a = \frac{1}{\mu}}(\alpha) && = 
 N^{\alpha \frac{\mu}{2}  }  \theta \left( 0  \leq  \alpha \leq   \frac{2}{\mu} \right)
\label{pialphatotaldeloc}
\end{eqnarray}

\section{ Conclusions }

\label{sec_conclusion}

In this paper, we have proposed to analyze
the delocalized non-ergodic phase of some random matrix models
via the Wigner-Weisskopf approximation for the dynamics from an initial site $j_0$.
 The main output of this approach is the inverse $\Gamma_{j_0}(N)$
 of the characteristic time to leave the state $j_0$
that provides some broadening $\Gamma_{j_0}(N) $ for the weights of the eigenvectors.
In this framework, the localized phase is recovered as the region where 
 the broadening $\Gamma_{j_0}(N) $
 is smaller in scaling than the level spacing $\Delta_{j_0}(N) $.
Here we have focused on the delocalized non-ergodic
 phase existing in the region of parameters
where the broadening $\Gamma_{j_0}(N) $ decays with $N$
 but is bigger in scaling than the level spacing $\Delta_{j_0}(N) $. 
Then the number of resonances grows only sub-extensively in $N$ as $\frac{\Gamma_{j_0}(N)}{\Delta_{j_0}(N)} $.
 For the Generalized-Rosenzweig-Potter (GRP) Matrix model, we have shown how
the present approach allows to recover the results obtained previously via other 
methods \cite{kravtsov_rosen,biroli_rosen,ossipov_rosen,ioffe}.
For the L\'evy generalization of the GRP model with $1<\mu<2$, 
we have obtained that the dynamics is governed by a stretched exponential
and we have computed the multifractal properties of eigenstates.

\end{document}